# The Local Emergence and Global Diffusion of Research Technologies: An Exploration of Patterns of Network Formation

Loet Leydesdorff [1] & Ismael Rafols [2]

**Abstract**

Grasping the fruits of "emerging technologies" is an objective of many government priority programs in a knowledge-based and globalizing economy. We use the publication records (in the *Science Citation Index*) of two emerging technologies to study the mechanisms of diffusion in the case of two innovation trajectories: small interference RNA (*siRNA*) and nano-crystalline solar cells (*NCSC*). Methods for analyzing and visualizing geographical and cognitive diffusion are specified as indicators of different dynamics. Geographical diffusion is illustrated with overlays to Google Maps; cognitive diffusion is mapped using an overlay to a map based on the ISI Subject Categories. The evolving geographical networks show both preferential attachment and small-world characteristics. The strength of preferential attachment decreases over time, while the network evolves into an oligopolistic control structure with small-world characteristics. The transition from disciplinary-oriented ("mode-1") to transfer-oriented ("mode-2") research is suggested as the crucial difference in explaining the different rates of diffusion between *siRNA* and *NCSC*.

[1] Amsterdam School of Communication Research (ASCoR), University of Amsterdam, Kloveniersburgwal 48, 1012 CX Amsterdam, The Netherlands; loet@leydesdorff.net
[2] SPRU –Science and Technology Policy Research, University of Sussex, Brighton, BN1 9QE, England; i.rafols@sussex.ac.uk

**Introduction**

New technologies can be expected to emerge in configurations of disciplines, skills, and potential users (demand, markets). Several sub-dynamics—production, invention, and distribution—converge at various junctions, and this process may involve feedback loops in different stages (cf. Kline & Rosenberg, 1986). With hindsight, technologies can be traced back to confluences among disciplines, to specific configurations of entrepreneurship and scholarship (e.g., Hughes, 1983 and 1987), the fruitfulness of new instrumentalities (Price, 1984; Shinn, 2005), inducement mechanisms, and focusing devices (Rosenberg, 1969). User participation and specification can be particularly salient in new software-based technologies (Von Hippel, 1988). In summary, focusing and need-determination (Teubal, 1979), the availability of relevant resources, and conducive configurations have been indicated as crucial to the emergence of new technologies.

Emergence, almost by definition, comes as a surprise. However, because of the long gestation periods of new technologies, early warning indicators of emerging technologies are high on the agenda of policy-makers. Important questions from a policy perspective arise: Can one still jump on the bandwagon once a new technology has taken off? Are there first-mover advantages? Such strategic questions have hitherto not been answered unambiguously by science, technology, and innovation research because the interactions among the various subdynamics (production, diffusion, markets) are nonlinear, probably technology-specific, and poorly understood (Debackere & Clarysse, 1998; Leydesdorff & Van den Besselaar, 1994; Vishwanat & Chen, 2006). Do new technologies emerge in critical configurations, "transdisciplinarily" in industrial or academic environments (Gibbons *et al*., 1994), in university-industry-government relations (Etzkowitz & Leydesdorff, 2000) or are they the fruits of longer-term endeavors in R&D that were to be shielded against short-term demand (Rosenberg & Nelson, 1994; Mowery *et al*., 2001).

From the perspective of hindsight, the record of government intervention programs in science and technology since the 1970s can be considered as a sequence of experiments in a world-wide laboratory. Studer and Chubin (1980), for example, evaluated the "War on Cancer" of US administrations in the 1970s. The results were disappointing; these authors stated (in a methodological appendix, at pp. 269 ff.) that since all the relevant dimensions were constantly changing, there could be no baseline for systematic evaluation. Leydesdorff & Van der Schaar (1987), for example, found in a scientometric comparison of wind and solar energy programs in the Netherlands during the 1980s that the solar-energy program had been successful while the wind-energy one failed even though the two programs were identical.

National contexts provide another source of variation (Lundvall, 1992; Nelson, 1993; cf. Carlsson, 2006). In a series of studies based on Germany, Weingart and his colleagues questioned whether processes of cognitive and technological development could be steered by government agencies otherwise than contextually (Van den Daele *et al*., 1977 and 1979). The self-organizing dynamics of markets and sciences develop at the global level, while governance can perhaps organize conditions for retention only at the local level. However, effective retention by reflexive policy makers assumes detailed knowledge of the relevant dynamics and their potential synergies (Beccattini *et al*., 2003; Bonaccorsi, 2008). Do science and technology develop together as in a tango (Price, 1984)? How can government and/or management set conditions? Is "transdisciplinarity" or "interdisciplinarity" a necessary (but probably not sufficient) condition for the emergence of new technologies at interfaces? (Wagner *et al*., 2011).

In summary, the literature suggests that (linear) "transfer" or (nonlinear) "translation" from one context to another are important mechanisms because knowledge and skills have to be packaged before they can move across interfaces or from one place to another (Latour, 1987, at p. 227). This packaging entails encoding. Codification of knowledge enables its globalization since knowledge can then be communicated (Foray, 2004; Leydesdorff, 2006). Emerging knowledge-based technologies "conquer the world" by processes of diffusion (Rogers, 1962). However, diffusion has not only a geographical, but also a cognitive component. The new technologies spread world-wide, but thereafter "the world" itself may have changed. For example, the use of the internet changed "distance" as a concept in communication. The meanings of the technologies themselves may also change on the fly. For example, when governments step in with large funding resources, scientists may be pressured to re-label their research interests in terms of the current fashion.

In the case of nanoscience and nanotechnology, for example, some of the original breakthroughs at interfaces between applied physics and chemistry during the 1990s—such as atomic force microscopy and carbon nanotubes—diffused rapidly into disciplinarily different areas, such as molecular biology, when large funds became available for the "nano" domain in the early 2000s. What previously had been studied at the "micro" scale was reoriented towards the "nano" scale (Bonaccorsi and Vargas, 2010). Of course, such a change of focus can be rationalized from the perspective of hindsight. In the laboratory, PIs have to combine a wealth of options at the bench with limited resources. Thus, the lab may be sensitive to funding programs, while the same development can from another perspective also be considered as the result of a self-organizing dynamics.

From this evolutionary perspective, external interventions can perhaps bend trajectories. In innovation studies, one has distinguished between gradual changes along technological trajectories and more radical change between technological regimes (e.g., Dosi, 1982; Freeman & Perez, 1988; Geels & Schot, 2007). What can be considered as bending a trajectory and what as qualitative change at the regime level? In our opinion, this research question is methodologically equivalent to asking for significance: did interventions make a difference? When did a new system emerge? Was a synergy involved? Such questions, however, can be addressed using measurements and statistics (Dolfsma & Leydesdorff, 2009).

For example, one can assume that a system reproduces itself along a trajectory and thus can be expected to exhibit the so-called Markov property, which states that without any other information, the best prediction of the next state of a system is its current one (e.g., Riba-Villanova & Leydesdorff, 2001). Triple-Helix indicators, for example, focus on the synergetic reduction of uncertainty in university-industry-government relations (Lengyel & Leydesdorff, in press; Leydesdorff & Sun, 2009). These various statistics have in common that one needs time-series of relatively complex data that can be analyzed in terms of analytically distinguishable dimensionalities (Frenken & Leydesdorff, 2000; Saviotti, 1988).

**Geographic *versus* cognitive diffusion**

On the basis of an extensive review of the literature, Boschma (2005; cf. Frenken *et al*., 2009 and 2010) argued that the geography of innovation can be investigated using five analytical dimensions and their associated proximities: geographical, cognitive, social, organizational, and institutional. According to these authors, previous studies of diffusion have focused mainly on one type of these possible "spaces." For example, classical studies of technology diffusion investigated diffusion in *social* networks, hence in terms of social and organizational distances (Rogers, 1962). Other studies have considered cognitive distances in terms of technological similarity of patents (e.g., Breschi *et al*., 2003) or the disciplinary similarity in publications (e.g., Kiss *et al*., 2010).

In this study, we focus on the two main dimensions of diffusion—the geographic one and the cognitive—and two recently emerging technologies: RNA interference and the application of nanocrystals in solar cells. Both these technologies are recent (within the last two decades), science-based, and specific. Both can also be considered as instrumentalities in Price's (1984) sense: they open up a range of new possibilities for research. A similar perspective is provided by Shinn's (e.g., 2005) "research technologies." Instrumentalities or research technologies are sufficiently specific and codified so that their labeling is stable in different (e.g., policy) contexts. Generic technologies such as "nanotechnology" or "synthetic biology" can more easily be relabeled or differently emphasized in different contexts. In these specific cases, however, expertise is needed to recognize a contribution as part of the research program. Thus, we run less the risk of a compositional fallacy in the sampling using a search string (Bonaccorsi & Vargas, 2010; cf. Heimeriks & Leydesdorff, in press).

We use recently developed "global maps of science" based on the ISI Subject Categories (Rafols *et al.*, 2010) to study the cognitive diffusion and interdisciplinarity of these developments over the last decade, and we compare the results with the geographical diffusion of these same publication patterns using Google Maps. How do cognitive and geographical patterns emerge? How could one define "globalization" in terms of a global regime in relation to these two diffusion patterns? Our expectation is that globalization at the regime level presumes *both* cognitive and geographical diffusion. Cognitive diffusion among the sciences is based primarily on codification into a useful "instrumentality" while one would expect geographical diffusion to be based on recognition of the economic or strategic value of the new developments.

Diesner & Carley (2010) recently reviewed the literature about innovation diffusion from the perspective of the new network sciences using a model developed by Milroy & Milroy (1985). This model builds on Granovetter's (1973) distinction between weak and strong ties. In the initial stages of an innovation, innovators are marginal (that is, weakly linked) to various groups, geographically and socially mobile, and behave "under-conformingly" to "deviantly" (Watts, 2007). Early adopters, however, are central and strongly tied members of the adopting groups (Rogers, 1962). Within such innovation networks, one can first expect preferential attachment to the innovating groups (Barabási, 2002; Barabási & Albert, 1999; Gay, 2010a and b). In a next stage, as dense and multiplex networks of researchers adopt, order is newly configured with a tendency to oligopolistic control in the research centers.

One can compare this development with the distinction in innovation studies between Schumpeter Mark I and II in the business environment (e.g., Freeman & Soete, 1997). In his older writings, Schumpeter (1912) theorized that entrepreneurs challenge existing firms through a process of "creative destruction," that is, by introducing new ideas and by innovating. In later contributions, Schumpeter (1943) paid more attention to the key role of large corporations as engines of economic growth by accumulating non-transferable knowledge in specific technological areas and markets. This view is sometimes referred to as "creative accumulation" leading increasingly to an oligopolistic control structure (Soete & Ter Weel, 1999; cf. Freeman & Perez, 1988).

We hypothesize that an oligopolistic control mechanism in the scientific context can be expected to function as cliques in a small world: each actor can reach other actors in just a few "handshakes" (Watts, 1998; Watts & Strogatz, 1999). The network then becomes dense and clustered. Such a change—from a network in which preferential attachment to the innovators dominates to an oligopolistic structure with small-world characteristics— would not be without policy implications. The stimulation of innovation in the pre-innovation phase requires that opportunities and resources be made available at the level of individual researchers (PIs) working in large research facilities or well-equipped universities (Heinze *et al*., 2007). Discretionary resources at the level of the relevant research groups and a lenient management style seem crucial in this initial phase.

In these initial stages of adoption, one can expect first-mover advantages because preferential attachment accumulates over time. Science policy initiatives, however, tend to react reflexively to developments in science and technology and therefore to lag. Once the resources are made available in a next stage, one can expect these to be absorbed by the leading centers in the oligopolistic structures that may have emerged in the meantime. From a policy perspective, the crucial task seems to develop policy instruments that enable local agents to jump on the globalizing bandwagons in time (Fujimura, 1988). Such policy instruments should, in our opinion, monitor the research process and be highly specific in terms of their knowledge content. In a later stage, migration policies and buying oneself into the new development(s) may be well-advised policy objectives. However, market forces may by that time already prevail.

**Descriptive statistics of the two emerging technologies**

*a. RNA Interference* (*siRNA*)

"RNA interference" originated as a research program in molecular biology with an article published in *Nature* in 1998 entitled "Potent and specific genetic interference by double-stranded RNA in *Caenorhabditis elegans*" (Fire *et al*., 1998). The two principal investigators—Craig Mello at the University of Massachusetts Medical School in Worcester, and Andrew Fire at Stanford University—received the Nobel Prize in medicine for this breakthrough in 2006.

RNA interference allows for control of the expression of DNA, for example, by inhibiting and silencing specific parts of the DNA. Although these mechanisms had been noticed before, the new research program focused on the production of specific small-scale molecules with potentially large (e.g., therapeutic) effects (Sung & Hopkins, 2006). These molecules have become known as *siRNA* (that is, small interference RNA) or *RNAi*. For this reason, we used the search string (ts=siRNA or ts=RNAi or ts="RNA interference" or ts="interference RNA")[3] in the *Science Citation Index-Expanded* at the Web-of-Science (WoS) for the years since 1998.[4] The search yielded a recall of 27,946 documents on February 10, 2010.

---

[3] Unlike "ti" for title-searches, "ts" searches also the abstracts and keywords of the papers under study.
[4] We experimented with different search strings. However, the string "RNA SAME interference" provides a few hundred false positives because "interference" is a common word in molecular biology. Sixty-seven records could be retrieved using this search string prior to 1998, but these are due mostly (65) to use of the letters RNAi as an abbreviation for something else.

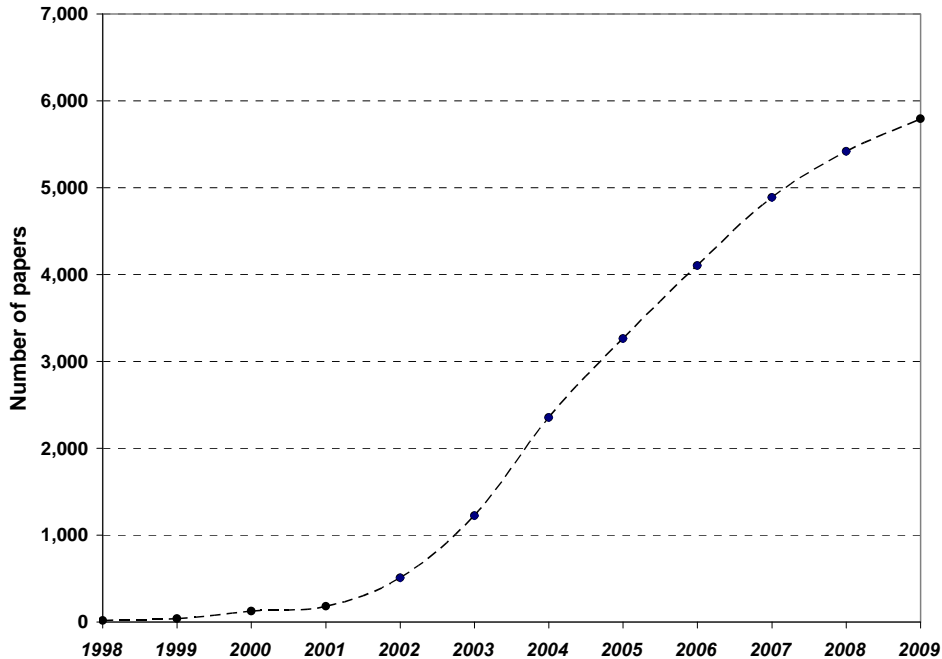

**Figure 1**: Number of papers per year retrieved with a topic search for "RNA interference" in the *Science Citation Index* 1998-2009.

Figure 1 provides the number of papers retrieved per year. (We used tape years instead of publication years so that the year 2009 was completed.)[5] The curve shows the well-known sigmoid shape. From 2002 onwards, the development enters the steep log-phase of the curve. During the last two years (2008 and 2009), the field seems to reach saturation. The other relevant parameters (numbers of authors, addresses, and cities) are highly correlated to the numbers of papers ($r > 0.99$; $p < 0.01$; $N = 12$). In summary, one can distinguish among three periods: an initial period from 1998 to 2002; a growth period from 2003 to 2007, and a saturation effect in the most recent period (2008-2009).

---

[5] The Web-of-Science interface allows for searching within the confines of a specific year (Jan. 1 - Dec. 31). These are the dates of entrance of the data into the machine and not their publication dates. Publication years can be searched with the tag "PY" in the search string.

Furthermore, an exploration of patent data taught us that indeed a commercially viable technology emerged from this research program. The initial period witnessed hardly any patenting activity; during the second period international patenting flourished at the *World International Patent Organization* (WIPO) in Geneva; and only in the last few years has patenting at the USPTO become substantial. Has increasing commercialization perhaps drawn the scientific development—as manifested in publications—into the saturation stage? (Shelton & Leydesdorff, in preparation).

*b. Nanocrystalline solar cells* (*NCSC*)

Solar cells themselves are an older technology based on the photovoltaic effect, that is, the conversion of light energy into electrical energy. The first solar cells were built in 1883 by Charles Fritts (who coated the semiconductor selenium with an extremely thin layer of gold to form the junctions) and the first US patent is from 1946.[6] Crucial for solar cells is their efficiency in turning light into electricity; research therefore focuses on developing highly efficient solar cells.

The possible development of solar cells based on nano-crystals became recognized as an option in the early 1990s. In their article entitled "Nanocrystalline Photoelectrochemical Cells - A New Concept in Photovoltaic Cells" (*J. of the Electrochemical Society* 139(11) (1992) 3136-3140), a team from the University of Southampton and the Weizmann Institute in Rehovot (Israel) stated the knowledge claim of this emerging technology as follows: "(S)emiconductor films with small crystal size normally exhibit prohibitively large recombination losses in photovoltaic cells. We show that porous nanocrystalline films of CdS and CdSe can be used as photoelectrodes in photoelectrochemical cells with relatively low recombination losses" (Hodes *et al.*, 1992). In order to place this development in the context of the "nano-revolution" (e.g., Guston, 2010), one should be aware that the first large funding programme started in the year 2000 (the US National Nanotechnology Initiative), precisely at the beginning of the growth phase shown in Figure 2. Nanocrystalline solar cells provide a "context of application" (Gibbons *et al.*, 1994) to the vision of nanotechnology.

---
[6] Source: http://en.wikipedia.org/wiki/Solar_cell; retrieved on July 22, 2010.

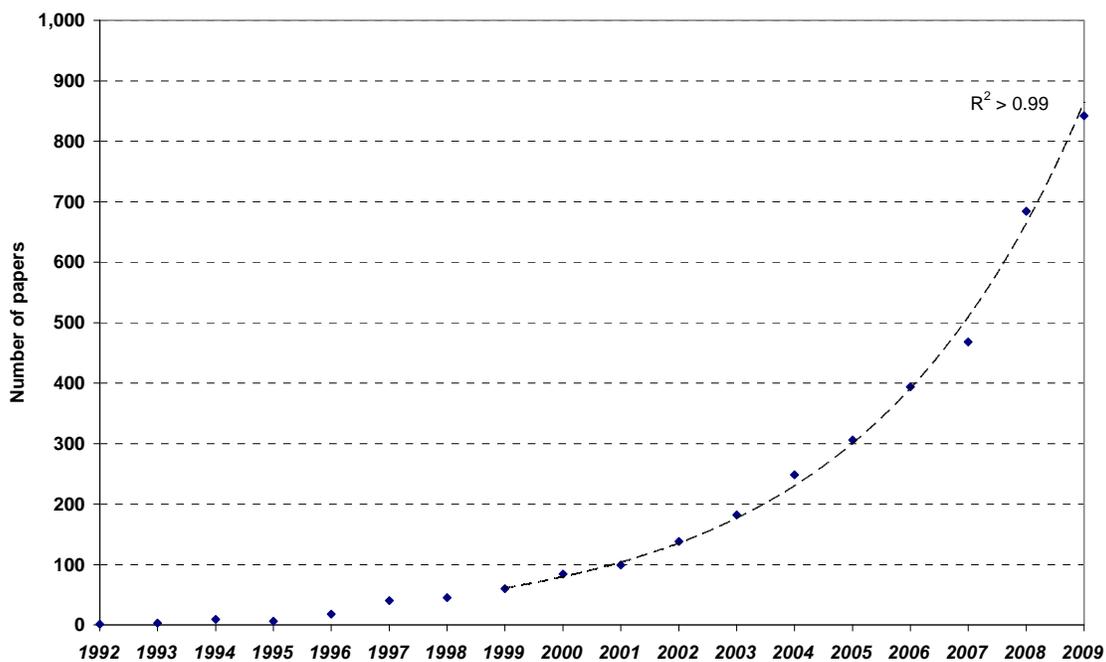

**Figure 2**: Number of papers about *NCSC* in the ISI database 1992-2009 ($N = 3,627$; February 3, 2010).

We used the following search string "ts=(nanocryst* or nano-cryst* or quantum-dot*) and ts=(solar-cell* or photovolt* or photo-volt*)" in the *Science Citation Index-*

*Expanded*. Figure 2 shows that the number of papers retrieved has increased exponentially over the last decade. Yet, this field has remained relatively small with fewer than 1,000 papers a year hitherto.

Furthermore, the technology is not at the stage of patenting. The same search string in the USPTO database retrieved only one single patent in 2005 (Nr. 6,849,798); there are 19 in the WIPO database (one in 1997, all the rest since 2005). However, searching with only the first part of this search string (that is, with "nanocryst\*" and its variants, but without the reference to photovoltaic cells) recalled approximately 400 patents in both databases. Thus, the specific combination of nano-crystals with solar cells in applications is technologically not yet a major source of patenting.

**Methods**

*Mapping and descriptive statistics*

Using the publication records for each year, we build on two methodological developments in bibliometrics to map the diffusion of emerging technologies both geographically and socio-cognitively. Leydesdorff & Persson (2010) developed (in anticipation of this project) a generic instrument for the mapping of institutional addresses (or cities) onto Google Maps. In this study, we extend this methodology by showing an animation of the spread of the new technologies as a global network. (The methodology is available at http://www.leydesdorff.net/maps.) The mapping technique is applied to the downloads for each consecutive year, and aggregated at the city level. City names which occur only once in a year are not included in the mapping. Cities which are connected by inter-institutional coauthorship relations are colored in red, and isolates in orange. The mapping thus allows visualizing not only the diffusion of a topic over the globe, but also the formation of the associated networks of collaboration.

Note that we did not aggregate or disaggregate at levels other than the city names. Cities can be part of metropolitan areas, but the literature about the problem of how to delineate metropolitan areas has hitherto remained inconclusive (Martin-Brelot *et al*., 2010; Pumain & Moriconi-Ebrard, 1997; Van Noorden, 2010). Thus, we used all address information as provided by the Web-of-Science, including the sometimes additional information of corresponding authors (Costas & Irribaren-Maestro, 2007). If this information is incomplete in the record, this problem often finds its origin in failing information in the underlying paper. We also did not decompose in terms of institutional addresses although that routine is available because this would in some cases overload the capacity of a Google map (Leydesdorff & Persson, 2010).

Geographic diffusion has an intuitive meaning: Google Maps allow us to zoom in or entertain a global view. Research-technologies, however, diffuse not only geographically but also across disciplinary borders (Chen, 2006; Chen *et al*., 2006; Morris, 2005). Using the ISI Subject Categories of the *Science Citation Index*, Rafols *et al*. (2010) developed a baseline for the visualization and measurement of diffusion across disciplines. Although the ISI classification system of journals contains a lot of error in the attributions, we argued that the randomness of this error allows us to use the aggregated citations among the categories for a comprehensive mapping of the sciences. The most recent map at the time of this research was based on the cosine-normalized cross-citation matrix among the 221 ISI Subject Categories attributed in the entire file of the *Science Citation Index* 2008 and *Social Science Citation Index* 2008 combined. (The map is available at http://www.leydesdorff.net/overlaytoolkit).[7]

*b. Testing social network properties*

The co-authorship networks among cities provide us with consecutive matrices for each year. These networks can be compared in terms of density, largest components, degree distribution, clustering coefficients, etc., using standard software for social network analysis such as Pajek and UCINet. Our main questions (in this geographical dimension) are: do patterns of diffusion change and how does this show in the development of various network parameters? What types of networks emerge? When does a network stabilize, and how?

We use the various indicators as proxies to operationalize heuristics: how can the diffusion of new technologies be characterized? For example, the slope in the degree distributions (of each year) can be used as indicators of preferential attachment (e.g., Barabási & Albert, 1999; Newman, 2004; Wagner & Leydesdorff, 2005). One can expect newcomers to attach themselves preferentially to leading centers. This dynamics can be expected to self-generate a configuration of relatively "small worlds" in which leading centers coauthor frequently with other leading centers. As a division of labor emerges, newcomers can be expected to attach themselves to these different centers in accordance to their disciplinary affiliations. In terms of organizational control, one thus can expect oligopolistic control by the centers to develop.

---

[7] See www.idr.gatech.edu/maps for an interactive version.

Following Watts & Strogatz (1998), Newman *et al*. (2006, at pp. 288f.) specified that a small world presumes that both the mean node-to-node distances (*d*) in an empirical network is comparable with that of a random graph ($d / d(rg) \sim 1$), and that the clustering coefficient (*CC*) is much larger in empirical networks than in the corresponding random graph (*CC(rg)*). Walsh' (1999) proximity ratio ($W = \frac{CC/CC(rg)}{d/d(rg)}$) would then be orders of magnitude larger than one for networks that contain the small-world property.[8]

Random graphs can be simulated on the basis of the parameters of an empirical network using the Erdős–Rényi model (available in Pajek). Newman *et al*. (2006, at pp. 286f.) has shown that an analytical approach to the problem yields $CC(rg) = z/n$, in which *z* is the average degree and *n* the number of nodes. In random graphs, the mean of the node-to-node distances scales with the logarithm of the number of nodes (*n*), and therefore it follows that $d(rg) = \log(n) / \log(z)$. The values thus estimated can be compared with the empirically simulated ones and a value for Walsh' proximity ratio can be determined. We use both this analytical approach and the one based on simulations.[9]

---

[8] Mark Newman (personal communication, 29 July 2010) noted that Watts & Strogatz (1998) defined the "small-world network" as one that shows both high clustering and the small-world effect. The latter is more narrowly defined (in physics) as a network where shortest distances scale with the log(*n*), with *n* as the number of nodes. We follow in this study the concept and terminology of Watts & Strogatz (1998).

Preferential attachment can be inferred by plotting the degree distribution of a network using the logarithms of the scales at both axes ("log-log plots"). A fit with a straight line provides a measure for preferential attachment because the power-law distribution follows $p_x = \alpha \cdot x^{-\gamma}$, and therefore: $\log(p_x) = -\gamma \cdot \log(x) + \log(\alpha)$.[10] Power-law distributions often have exponent values of $\gamma$ between 2 and 3. However, Powell *et al.* (2005, at p. 1153) reported values as low as 1.1 for empirical networks of collaboration in the life sciences.[11] These lower values in social networks—when compared with biological networks or networks among molecules—were attributed to dynamics other than preferential attachment operating at the same time.[12]

*c. Diffusion in the cognitive dimension*

Maps showing cognitive distances among journal categories are not relational, but projections in two dimensions of a cosine-based (multidimensional) vector space. A spring-embedded algorithm can be used for the reduction of complexity that allows for the visualization (Kamada & Kawai, 1989). However, distances in this two-dimensional projection cannot be made the subject of social network analysis without a reflexive translation since the cosine is not a relational but a similarity measure.

---

[9] In principle, the simulated values are more precise given a very large number of runs.
[10] Preferential attachment is a theoretical model of network dynamics that leads to the expectation of power-law distributions, but power-law distributions can be generated in a number of different ways (e.g., Katz, 2000).
[11] Jeong *et al.* (2003, at p. 570) found values smaller than unity when studying coauthorship networks in the neurosciences. However, these authors used cumulative degree distributions and therefore the values should be incremented with unity (i.e., $\gamma = \alpha + 1$).
[12] A collaborative tie was defined by Powell *et al.* (2005) as any contractual arrangement to exchange or pool resources between a dedicated biotechnology firm and one or more partner organizations.

Rafols & Meyer (2010) suggested that one could use the diversity measure proposed by Rao (1982) and Stirling (2007) to measure diversity in such configurations. (This measure is sometimes also called "quadratic entropy"; e.g., Izsáki & Papp, 1995). Rao-Stirling diversity (*D*) is defined as follows:

$$D = \sum_{ij(i \neq j)} p_i \cdot p_j \cdot d_{ij} \tag{1}$$

in which $p_i$ represents the relative frequency of the occurrence of each category *i,* and $d_{ij}$ the distance in the network between each two nodes *i* and *j*. Unlike traditional measures of diversity (such as Shannon entropy or the Herfindahl index), diversity is thus measured not only between categories, but also in terms of the cognitive distances between them, that is, by using an ecological perspective (Ricotta & Szeidl, 2006). Diversity can be increased both among categories and by spreading in terms of relevant categories across the network.

In a number of studies, we tested this measure as an indicator of interdisciplinarity in bibliometric datasets, but using essentially a static design (Leydesdorff & Rafols, 2011; Porter & Rafols, 2009; Rafols & Meyer, 2010). In this study, we explore whether the spread across disciplines during the diffusion process can also be measured as an increase in the diversity. Is this increase continuous or does it change over time?

*d. Measures of globalization*

The diversity measure noted above is generic and can therefore be applied to any type of categories; for example, the city names studied in the geographic dimension. In this case, distances ($d_{ij}$) can be expressed, for example, in kilometers (e.g., at http://www.movable-type.co.uk/scripts/latlong.html). From this perspective, the term ($p_i \cdot p_j$) in Equation 1 can be considered as the *expected* relationship between two cities *i* and *j*. One can analogously define a measure of globalization in the *observed* coauthorship relationships $p_{ij}$ as follows:

$$C = \sum_{ij(i \neq j)} p_{ij} \cdot d_{ij} \qquad (2)$$

This measure accounts for both the relative frequencies of collaborations ($p_{ij}$) and the distance ($d_{ij}$) between the cities. In other words, *C* measures the average distance among cities related in the network. *C* and *D* can also be compared as observed versus expected values of globalization of co-authorship relations (Wagner, 2008; Wagner & Leydesdorff, 2005).

Note that the value of the diversity (*D*) is independent of whether or not a network exists among (some of) the nodes. Only the distribution of the events (in this case, papers) over the nodes and their respective distances is taken into account.[13] However, observe coauthorships (*C*) provide us with a network indicator of the same dimensionality as the Rao-Stirling diversity indicator (*D*). The ratio between these two measured can be considered as a measure of coherence (Rafols *et al.*, in preparation.)

---
[13] One can also ask for the diversity among the nodes which are connected in the network as a subset.

**Results**

*Mapping the Geographic Diffusion*

The co-authorship networks for the two technologies are different in terms of the development of their connectedness. The development of *siRNA* shows an increasing connectedness within the network to approximately 100%. In 2009, more than 95% of the locations with more than a single paper are related through network connections to the largest component. In the case of *NCSC* only 78.6% of the 271 cities in which more than a single paper is produced are networked. The number of institutional addresses/paper in 2009 is 2.77 for *siRNA* and 2.18 for *NCSC*. Thus, the growth is not only an order of magnitude larger in the case of *siRNA*, but also the evolution of the network is slower in the case of *NCSC*.

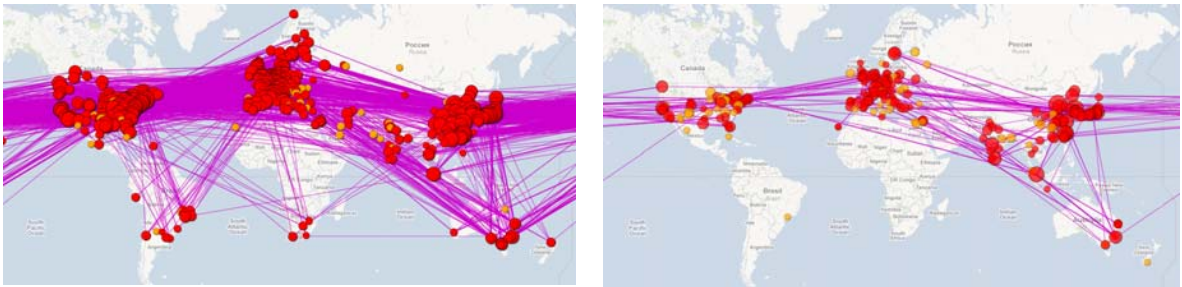

**Figure 3**: 2009-networks of intercity collaboration in *siRNA* and *NCSC*, respectively.

As an example, Figure 3 shows the two network configurations graphically as overlays to Google Maps. The maps for all years can be accessed at http://www.leydesdorff.net/et/sirna.html and http://www.leydesdorff.net/et/nanocr.html, respectively. The development of the networks can additionally be animated at these sites.

In both networks the clustering coefficients increase and the average distances decrease, but both these tendencies are stronger and begin earlier in the case of *siRNA* than in that of *NCSC*. The network of *siRNA* is almost saturated since the mid-2000s, while that in *NCSC* has lagged behind in this respect since the first half of the 2000s. However, the density has stabilized between one and two percent in both networks since 2001: although the vast majority of centers are connected in one way or another, connections remain relatively scarce when compared with possible relations.

The development of the average distance (not shown here) goes through a maximum in the case of *siRNA* in 2002 but in the case of *NCSC* only in 2006. The development of these and various other parameters suggests that the pace of development changed in the case of *siRNA* after 2002: the network of collaborations becomes rapidly saturated at the global level. The network developed slower and more gradually in the case of *NCSC* during the early 2000s. However, the eventual results in 2009 are not so different, although they developed at a different pace.

*Cognitive Diversity*

Let us proceed with a description of diffusion in the cognitive dimension. For this purpose, we took the maps of journal categories that were created by Rafols *et al.* (2010) as a baseline and overlaid these maps with the contributions in each year. When papers are published in journals with more than a single category attribution, they are counted in both categories. Thus, one can generate an overlay for the sets for each year and animate the diffusion across the cognitive landscape. The animations are available at http://www.leydesdorff.net/et/diversity/siRNA.wmv and http://www.leydesdorff.net/et/diversity/nanocr.wmv, respectively. Figures 4 and 5 show the results of the diffusion processes for the two technologies in 2009.

Whereas *siRNA* (Figure 4) has diffused from its original position in the category of Biomedical Sciences, first to Clinical Medicine and Infectious Diseases, by 2003 this diffusion also reached Chemistry and went from there to Agriculture, Ecology and the Materials Sciences. *NCSC* (Figure 5) diffused initially from Materials Science to Chemistry and the Enviromental Sciences, but the Rao-Stirling diversity in this area no longer increased after 1997. Animations of the diffusion patters in the two domains over the periods under study are available at http://www.leydesdorff.net/et/diversity/sirna.wmv and http://www.leydesdorff.net/et/diversity/nanocr.wmv, respectively.

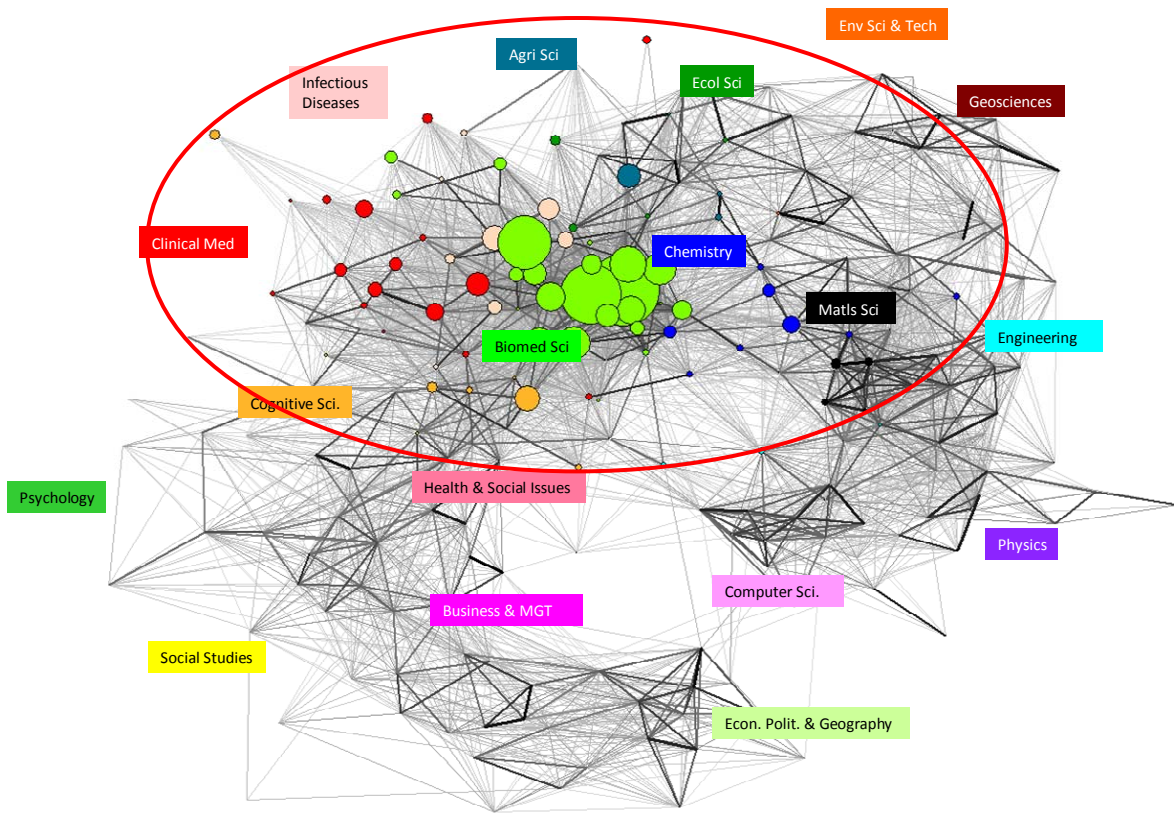

**Figure 4**: Diffusion of papers about *siRNA* across the cognitive map (in 2009).

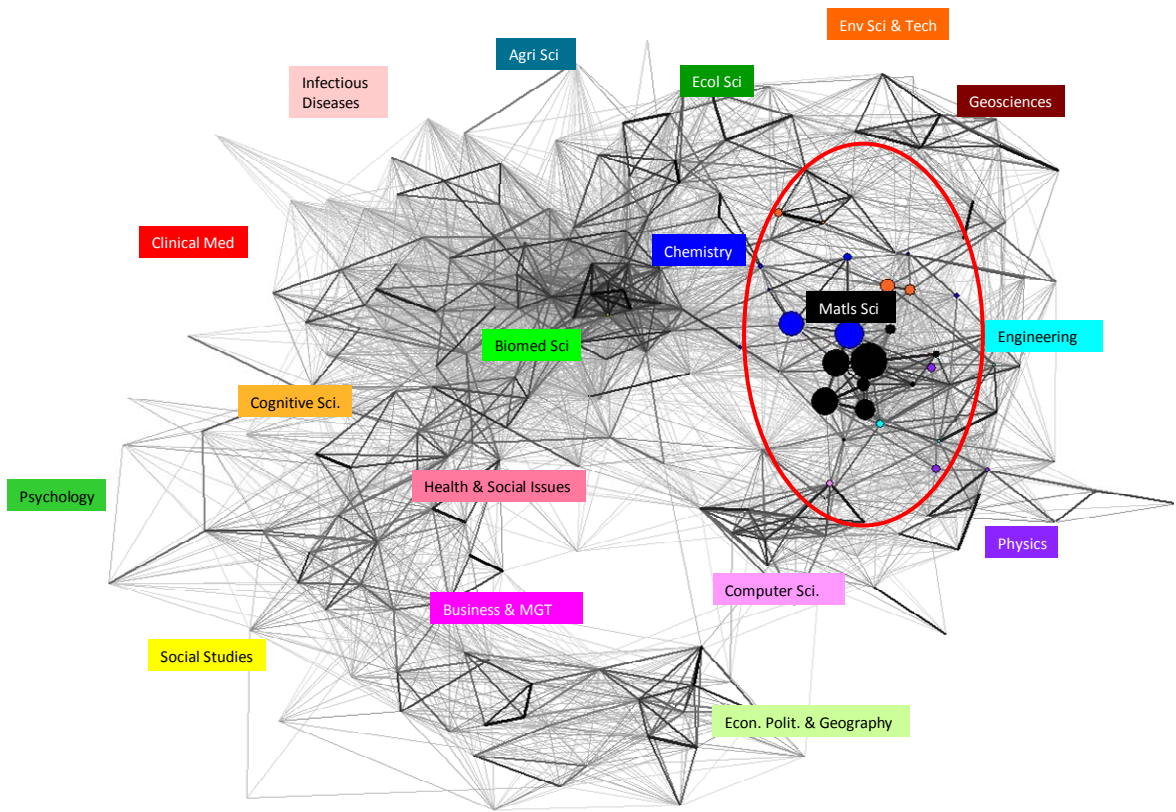

**Figure 5**: Diffusion of papers about *NCSC* across the cognitive map (in 2009).

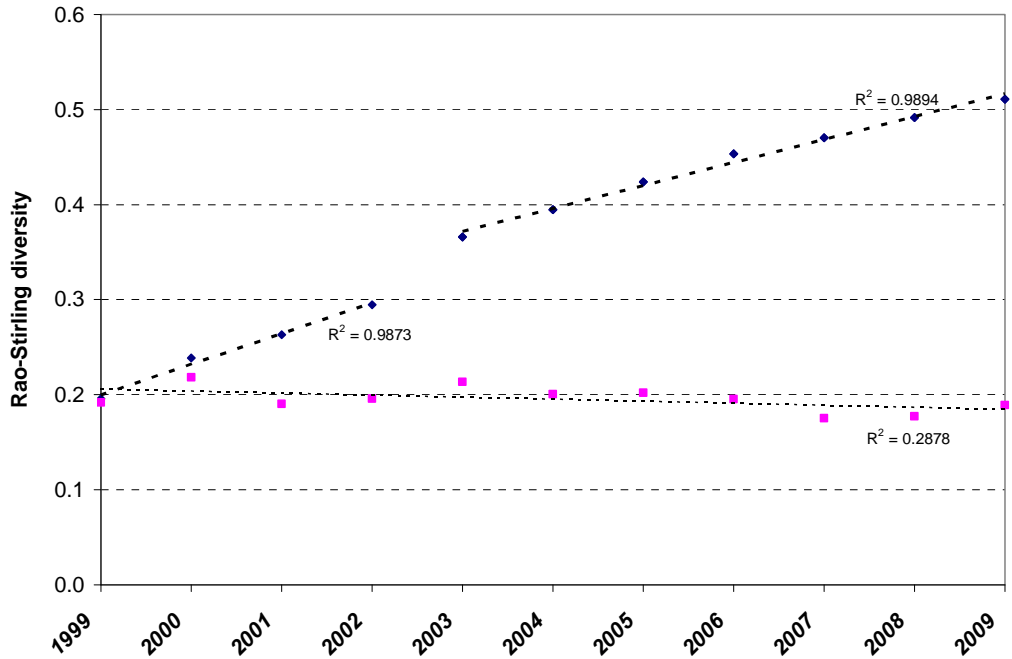

**Figure 6**: Rao-Stirling diversity in the ISI Subject Categories 1999-2009 for *siRNA* (♦) and *NCSC* (■), respectively.

Figure 6 shows that Rao-Stirling diversity in terms of these disciplinarily organized subject categories did not increase during the last decade for *NCSC*, but continued to increase for *siRNA*. Between 2002 and 2003, this increase (in other words, the slope of the curves) reached a different level. As we shall see later in this study, this difference between the two research technologies in terms of cognitive diversity is more significant than any of the differences in geographical diffusion patterns.

*Observed and expected coauthorship relations*

One can apply the Rao-Stirling diversity measure also to global spread of the developments by using geographical distances as input to the distance matrix and then compare with observed coauthorship relations. In this case, the geographical distances ($d_{ij}$) can be provided with a dimensionality (e.g., kilometers; at http://www.movable-type.co.uk/scripts/latlong.html).

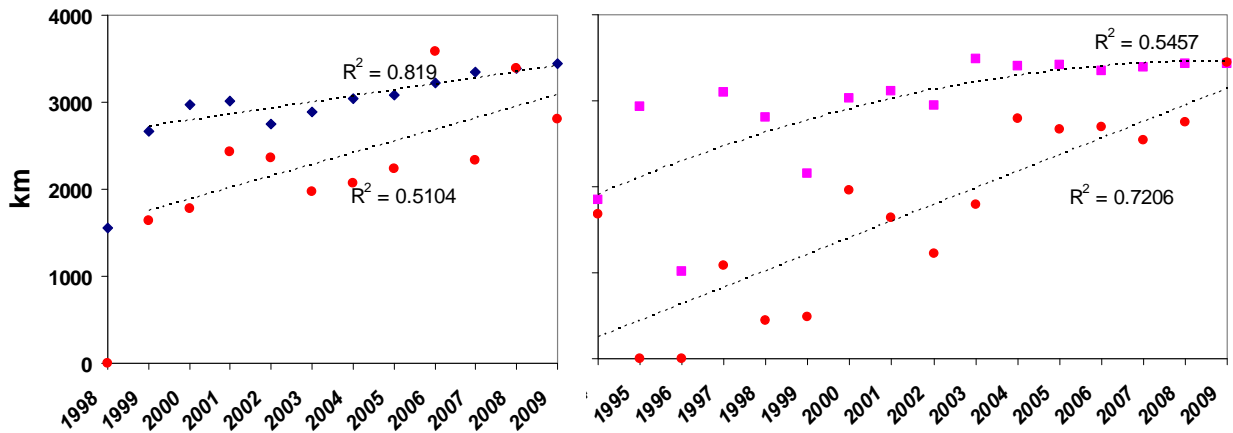

**Figure 7**: The development of global collaboration (●) and geographical diversity for *siRNA* (♦; left) and *NCSC* (■; right), respectively.

Figures 7 shows the evolution of this global collaboration and diversity as indices of globalization for *siRNA* (on the left) and *NCSC* (on the right), respectively. In both cases, the geographical spread is more stable than the collaboration measure. Since 2004, however, the relative collaboration rate (that is, observed versus expected collaboration) is high (> 80%) in both technologies.

*c. Preferential attachment and small worlds*

Can relevant differences between the two technologies be indicated using the various parameters available from social network analysis such as clustering coefficients, preferential attachment, and small worlds? Among these various parameters, the emergence of small worlds was hypothesized by Gay (2010a and b) as specifically important in an evolutionary process because the emergence of a small world indicates a phase transition. A phase transition can be indicated by a change in dynamics.

Is the original dynamics of the diffusion of an invention among individual scholars replaced with the dynamics of organized knowledge production and control in networks with a dynamics at the supra-individual level? (Whitley, 1984) Can a dynamics of transition equivalent to the one between Schumpeter Mark I (entrepreneurial initiative to innovation) and II (large corporations lead the innovations) thus perhaps be indicated? (Freeman & Soete, 1997)

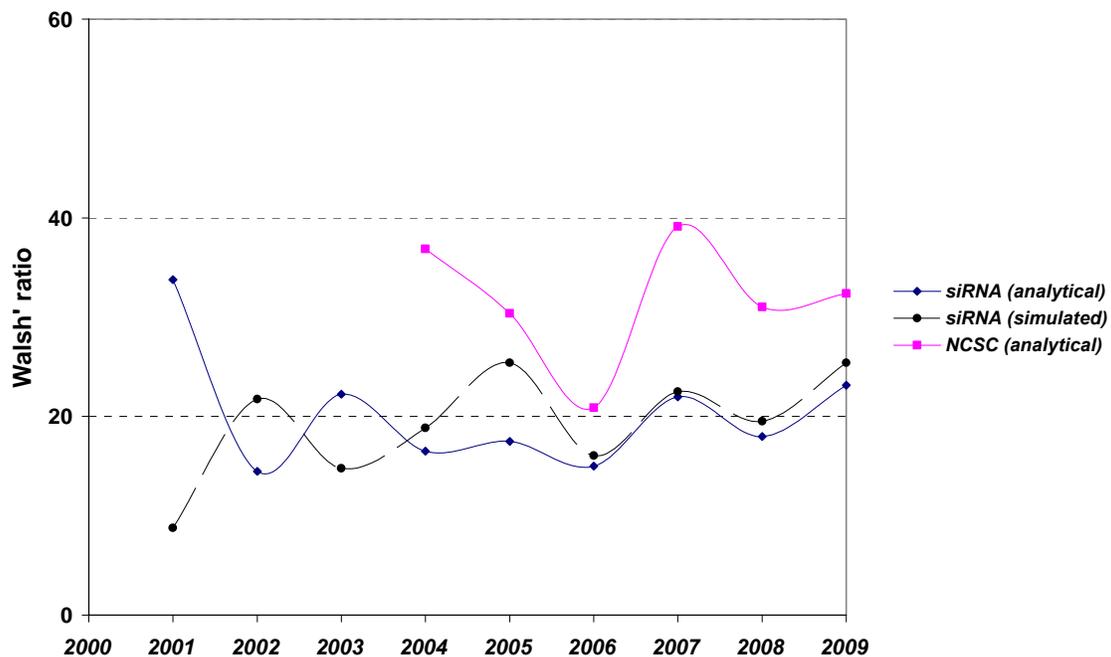

**Figure 8**: Testing the hypotheses of small-world networks using address information in *siRNA* (♦) and *NCSC* (■) research, respectively.

As noted, Walsh' proximity ratio ($W = \frac{CC/CC(rg)}{d/d(rg)}$) can be used to test whether small-world networks emerged in the two research technologies under study. Figure 8 shows that both the analytical solution and the simulation-based one indicate a level of approximately $W = 20$ for the *siRNA* set from 2002 onwards. Before 2001, the development of $W$ is very erratic because of low clustering coefficients or values of the average degree ($z$) close to one (and therefore, log($z$) as used in calculating the denominator of $d(rg)$ close to zero).[14] However, on average the simulated values are of the same order of magnitude as the analytical values from 2004 onwards. The two research technologies are not significantly different in these terms after this date.

In other words, a substantial and relatively stable small-world network effect is indicated in both technologies, but somewhat later in *NCSC* than in *siRNA*. The values of Walsh' proximity ratios in the later years are considerably larger than one, but only by a single order of magnitude. Note that the use of the terminology "small worlds" in this context of relations among cities is a bit metaphorical since cities themselves are not the agents of collaboration. Cities can be considered as centers of institutional agency. Thus, one can say that some cities act as centers that increasingly take over the dynamics from the agency of individual scholars. Authors in cities at the periphery of the network are related through these central cities. In summary, such a change in the dynamics is indicated in both technologies albeit at different moments of time.

---

[14] We did not include the simulation-based values for *NCSC* in Figure 10 because these contain an outlier above hundred in 2009.

Might perhaps the mechanisms of preferential attachment be affected differently before and after this transition? Among the various descriptive statistics of the networks, the degree distribution can be used to test another a theoretically informed expectation, namely, preferential attachment (Barabási & Albert, 1999).

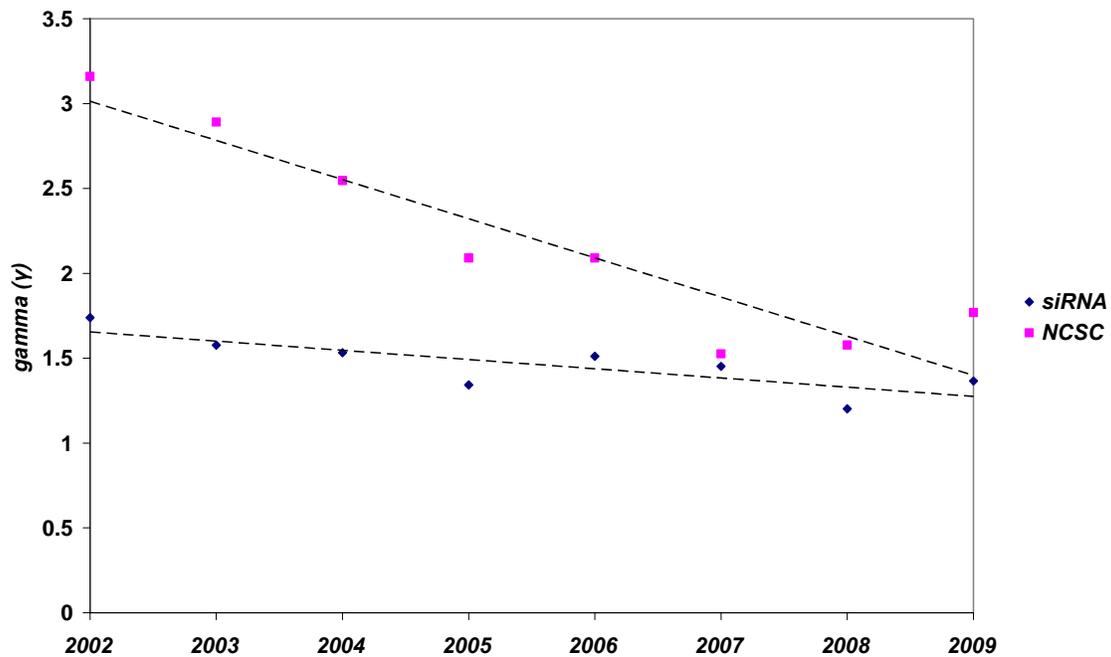

**Figure 9**: Exponents of the degree distributions 2002-2008 in *siRNA* (♦) and *NCSC* (■) research, respectively.

Figure 9 shows the development of the exponents (– γ) in the log-log plot of the degree distribution in the networks of cities with publications in *siRNA* and *NCSC*, respectively. The underlying log-log plots show reasonably good fits ($r^2 \approx 0.8$) after correction for the hooks of the respective distributions.[15] Although the preferential attachment patterns decrease over time, a transition is not indicated. Both before and after the emergence of the small-world effect, the networks are scale-free. As noted, Powell *et al*. (2005, at p. 1153) hypothesized that social dynamics other than preferential attachment can lead to decreases in the values of the exponent.

---

[15] The hooks in the log-log plots are caused by new entrants to the networks with degrees of unity or only a few. As usual, these points were not included in the regression.

Diesner & Carley (2010) noted that the original innovators may be different from the first adopters. Whereas the latter can be expected to come from the best-organized and integrated groups (Rogers, 1962), the innovators are relatively isolated (Watts, 2007). In the case of *NCSC*, for example, the original paper originated from institutional addresses in Rehovot, Israel and Southampton, England, while the leading centers have become the Swiss Federal Institute of Technology (EPFL) in Lausanne (Switzerland) and a group of researchers at the Imperial College in London.

|   | *siRNA*   | > 100 links | *NCSC*   | central cities |
|---|-----------|-------------|----------|----------------|
| 1 | Boston    | 141 | Lausanne | 17 |
| 2 | Bethesda  | 131 | London   | 12 |
| 3 | London    | 127 | Golden   | 9 |
| 4 | Houston   | 124 | Seoul    | 9 |
| 5 | Cambridge | 109 | Shanghai | 9 |
| 6 | Heidelberg| 106 | Taejon   | 9 |
| 7 | Berlin    | 103 | seven cities with eight links | |

**Table 1**: Central cities in the two research technologies (2008).

In 2008, for example, the top-13 cities involved in *NCSC* maintain eight of more links and form a dense network (Table 1). The EPFL in Lausanne functioned as the incubator of the original invention of this technology in the early years. Their coauthor link to the Imperial College in London, for example, was established in 1997. The presence of Asian cities (Seoul, Shanghai) in this technology is also to be noted, while the *siRNA* technology is firmly in the hands of advanced and highly-industrial nations.

**Conclusions and discussion**

Using the postal addresses in the bylines and the disciplinary affiliations of the journals in which publications appeared we traced the diffusion of two research technologies both geographically and in the cognitive domain. The geographical diffusion patterns show that while *siRNA* has indeed been fully globalized, publications about *NCSC* are diffusing at a slower pace. However, *siRNA* seems to have entered a phase of commercialization and perhaps more recently saturation of the academic research trajectory, while academic knowledge production in *NCSC* does not yet seem to be in that phase.[16]

What caused the difference between the two diffusion patterns? We found one clue that explains this difference: the development of Rao-Stirling diversity in the cognitive domain (Figure 6). After an initial period (1998-2002), *siRNA* research went into a different gear: the research technology became transferable to other areas of research than the one from which it originated and a division of labor became established at the global level. This research system became almost completely networked. The transferability of the research technology to other cognitive domains seems to have been the crucial condition for this further development.

---

[16] However, a conference on "Dye Solar Cell Industrialisation" in 2007 formulated on the website at http://blogs.epfl.ch/dyesolarcell: "While the conference does not stress academic research it does include a session on modelling and next generation DSC devices and designs (such as tandem devices)."

Research in *NCSC* also experienced an increase in Rao-Stirling diversity during the 1990s, but this came to a halt during the 2000s. This research technology grew further *within* its disciplinary structures. The centers of production of articles also dominate the network of collaborating centers, but the number of coauthored articles between these centers are an order of magnitude smaller than in the network between the leading centers in *siRNA* research. The size of the operation is smaller, globalization is more modest, and transfer to disciplinary affiliations other than those already involved has stagnated.

In terms of the geographical network of cities, we found small-world network effects in both cases, but the function of the network among centers may be different in the two technologies. In *NCSC*, the prevailing pattern has remained mainly disciplinary, while in *siRNA* the pattern became interdisciplinary. Perhaps, this is partly inherent to the differences between new developments in the life sciences (*siRNA*) and the material sciences (*NCSC*). The difference between the two cases may depend on the fact that *siRNA* is a research technology potentially useful to many applications, whereas *NCSC* is already applied in a specific applicational context, and therefore less apt to migrate to other cognitive domains.

The networked system in *siRNA* is saturated: more peripheral centers are connected to one another through the laboratories in the center. Because of the division of labor in this oligopolistic structure, peripheral centers can compete *within* their disciplinary compartments. The continuously driving mechanism of preferential attachment therefore is moderated. Such crystallization into an oligopolistic control structure seems not yet to be the case to the same extent in *NCSC*. The dominant structure has remained disciplinary. Yet, small-world network effects were retrievable. In a more disciplinarily organized structure, however, one does not need so urgently to connect to the center to access resources since academic ("mode 1") work can also be performed at smaller scales. What counts (in "mode 1" research) is publishing—more than networking (in "mode 2" research; Gibbons *et al.*, 1994).

Note that we have used the lens of publications for this analysis. (The development of *NCSC* does not yet spill over into patents, whereas *siRNA* is beginning to do so.) Five analytical dimensions were distinguished above—geo, cognitive, social, organizational, and institutional—and our (neo-)evolutionary model can further be specified in terms of the dynamics among them: in the initial stages of the (cognitive) inventions, diffusion proliferates as variation from the center(s). A niche is shaped and further developed along a historical trajectory. The social network mechanisms, however, provide their own dynamics of preferential attachment and potential lock-in into a small world. These social and organizational mechanisms generate a network dynamics among centers which can take over control (Hughes, 1987). In this regime phase, geographical distance becomes less of a barrier because the centers are increasingly connected at a global level.

From a historical perspective, the differences in the development between the two research technologies may seem incidental. Our conclusions rely on the assumption that the search strings were appropriate. In the cases presented here, we took care that the

keywords were unambiguous and specific.[17] Nevertheless, the possibility remains that the search terms were not accurate enough. Application-oriented publications (for example, in trade journals) Our results, however, suggest that the transition around 2003 from an explorative network to a transfer-oriented network in the case of *siRNA* can be considered as a phase transition (Bonaccorsi, 2008; Bradshaw & Lienert, 1991; Frenken, 2005). After this phase transition, the research technology itself became application-oriented. Our analysis thus suggests focusing the policy efforts on facilitating this transition.

**Acknowledgment**
We are grateful to Brigitte Gay, Staša Milojevic, and Koen Frenken for suggestions and comments on previous drafts.

---

[17] Using the search string "thin-film and (solar or photovoltaic)", Guo *et al.* (2009) found 1,659 records from the *Science Citation Index Expanded* (during the period 2001 to mid-2008), but with a very different profile in terms of main cities. Note that "nanocrystalline" is a subset of thin-film solar cells, including also micro-crystalline solar cells.